\documentclass[runningheads]{svmult}
\usepackage{graphicx}  
\usepackage{physprbb}  
\def \aap #1 #2 {{Astron. Astrophys.\/} {\bf #1}, #2}
\def \aas #1 #2 {{Astron. Astrophys. Suppl. Ser.\/} {\bf #1}, #2}
\def \ar #1 #2 {{Astron. Rep.\/} {\bf #1}, #2}
\def \araap #1 #2 {{Ann. Rev. Astron. Astrophys.\/} {\bf #1}, #2}
\def \apj #1 #2 {{Astrophys. J.\/} {\bf #1}, #2}
\def \apjl #1 #2 {{Astrophys.. J. Lett.\/} {\bf #1}, #2}
\def \baas #1 #2 {{Bull. Am. Astron. Soc.\/} {\bf #1}, #2}
\def \basi #1 #2 {{Bull. Astron. Soc. India\/} {#1}: #2}
\def \cap #1 #2 {{Comm. Astrophys.\/} {\bf #1}, #2}
\def \iauc #1 {{IAUC\/} {#1}} 
\def \iaus #1 #2 {{IAU Symp. 110: VLBI \& Comp. Rad. Sour.\/} {\bf #1}, #2}
\def \gcn #1 {{GCN\/} {#1}} 
\def \mnras #1 #2 {{Mon. Not. R. Astr. Soc.\/} {\bf #1}, #2}
\def \nat #1 #2 {{Nature\/} {\bf #1}, #2}
\def \newa #1 #2 {{New Astron.\/} {\bf #1}, #2}
\def \nyasa #1 #2 {{NY Acad. Sci. Ann.\/} {\bf #1}, #2}
\def \pasa #1 #2 {{Pub. Astron. Soc. Australia\/} {#1}: #2}
\def \pasp #1 #2 {{Pub. Astron. Soc. Pacific\/} {\bf #1}, #2}
\def \sci #1 #2 {{Science\/} {#1}: #2}
\def \TX17 #1 #2 {{Seventeenth TX Symp. Rel. Astrophys. Cosmol.\/} {\bf #1}, #2}
\def\B{Beppo-Sax~}
\def\CO{Compton Gamma-Ray Observatory~}
\def\I{IRAS~}
\def\J{James Clerk Maxwell Telescope~}
\newcommand{\kms}{km~s$^{-1}$~}
\newcommand{\ie}{i.e.,~}
\newcommand{\etal}{et al.}
\newcommand\arcdeg{\mbox{$^\circ$}}%
\newcommand\arcmin{\mbox{$^\prime$}}%
\newcommand\arcsec{\mbox{$^{\prime\prime}$}}%
\newcommand\fs{\mbox{$.\!\!^{\mathrm s}$}}%
\newcommand\farcs{\mbox{$.\!\!^{\prime\prime}$}}%
\def\gtrsim{\mathrel{\hbox{\rlap{\hbox{\lower4pt\hbox{$\sim$}}}\hbox{$>$}}}}

\begin{document}
\title*{Radio Observations of GRB Afterglows}
\toctitle{Radio Observations of GRB Afterglows}
%
%
\titlerunning{GRB Radio Afterglows}
%
\author{Kurt W. Weiler\inst{1}
\and Nino Panagia\inst{2}
\and Marcos J. Montes\inst{3}
}
\authorrunning{Kurt W. Weiler et al.}
\institute{Naval Research Laboratory, Code 7213, Washington, DC 20375-5320, USA 
\and Space Telescope Science Institute, 3700 San Martin Drive, Baltimore, MD 21218, USA \& Astrophysics Division, Space Science Department of the European Space Agency \and Naval Research Laboratory, Code 7212, Washington, DC 20375-5320, USA}

\maketitle              

\begin{abstract}
Since 1997 the afterglow of $\gamma$-ray bursting sources (GRBs) has occasionally been detected in the radio, as well in other wavelengths bands.   In particular, the interesting and unusual $\gamma$-ray burst GRB980425, thought to be related to the radio supernova SN1998bw, is a possible link between the two classes of objects.  Analyzing the extensive radio emission data available for SN1998bw, one can describe its time evolution within the well established framework available for the analysis of radio emission from supernovae.  This then allows relatively detailed  description of a number of physical properties of the object.  The radio emission can best be explained as the interaction of a mildly relativistic ($\Gamma \sim1.6$) shock with a dense preexplosion stellar wind-established circumstellar medium (CSM) that is highly structured both azimuthally, in clumps or filaments, and radially, with observed density enhancements.  Because of its unusual characteristics for a Type Ib/c supernova, the relation of SN1998bw to GRB980425 is strengthened and suggests that at least some classes of GRBs originate in supernova (SN) massive star explosions.  Thus, employing the formalism for describing the radio emission from supernovae (SNe) and following the link through SN1998bw/GRB980425, it is possible to model the gross properties of the radio and optical/infrared (OIR) emission from the half-dozen GRBs with extensive radio observations.  From this we conclude that at least some members of the ``slow-soft'' class of GRBs can be attributed to the explosion of a massive star in a dense, highly structured CSM that was presumably established by the preexplosion stellar system.
\end{abstract}

\section{SN1998bw/GRB980425}

The suggestion of an association of the Type Ib/c SN1998bw with the $\gamma$-ray burst GRB980425 may provide evidence for another new phenomenon generated by SNe -- at least some types of GRBs may originate in some types of SN explosions.  Because SN1998bw/GRB980425 is by far the nearest and best studied of the $\gamma$-ray bursters, it is worthwhile to examine its radio emission in detail before proceeding to the discussion of the radio emission from other GRBs.

\subsection{Background}

While generally accepted that most GRBs are extremely distant and energetic (see, e.g., \cite{Paczynski86,Goodman86}), the discovery of GRB980425 \cite{Soffitta98} on 1998 April 25.90915 and its possible association with a bright supernova, SN1998bw at RA(J2000) = $19^h 35^m 03\fs31$, Dec(J2000) = $-52\arcdeg 50\arcmin 44\farcs7$ \cite{Tinney98}, in the relatively nearby spiral galaxy ESO~184-G82 at $z = 0.0085$ (distance $\sim40$ Mpc for $H_0 = 65$ \kms Mpc$^{-1}$) \cite{Galama98a-KW,Galama98b-KW,Galama99,Lidman98,Sadler98,Tinney98,Woosley99}, has introduced the possibility of a SN origin for at least some types of GRBs.  The estimated explosion date of SN1998bw in 1998 between April 21 -- 27 \cite{Sadler98} corresponds rather well with the time of the GRB980425 outburst.  Iwamoto \etal \cite{Iwamoto98-KW} felt that they could restrict the core collapse date for SN1998bw even more from hydrodynamical modeling of exploding C + O stars and, assuming that the SN1998bw optical light curve is energized by $^{56}$Ni decay as in Type Ia SNe, they then placed the coincidence between the core collapse of SN1998bw to within $+0.7$/$-2$ days of the outburst detection of GRB980425.

Classified initially as an SN optical Type Ib \cite{Sadler98}, then Type Ic \cite{Patat98}, then peculiar Type Ic \cite{Filippenko98,Kay98}, then later, at an age of ~300 - 400 days, again as a Type Ib \cite{Patat99}, SN1998bw presents a number of optical spectral peculiarities that strengthen the suspicion that it may be the counterpart of the $\gamma$-ray burst.

When the more precise \B NFI was pointed at the \B error box 10 hours after the detection of GRB980425, two X-ray sources were present \cite{Pian99}.  One of these, named S1 by Pian \etal \cite{Pian99}, was coincident with the position of SN 1998bw and declined slowly between 1998 April and 1998 November.  The second X-ray source, S2, that was $\sim4\arcmin$ from the position of SN 1998bw, was not (or at best only marginally with a $<3\sigma$ possible detection six days after the initial detection) detectable again in follow up observations in 1998 April, May, and November \cite{Pian99,Pian00-KW}. 

However, although concern remains that the Pian \etal \cite{Pian99,Pian00-KW} X-ray source, S2, might have been the brief afterglow from GRB980425 rather than the Pian source S1 associated with SN1998bw, Pian \etal \cite{Pian00-KW} concluded that S1 has a ``high probability'' of being associated with GRB980425 and that S2 is more likely a variable field source.

\subsection{Radio Emission}

The radio emission from SN 1998bw reached an unusually high 6 cm spectral luminosity at peak of $\sim6.7 \times 10^{28}$ erg s$^{-1}$ Hz$^{-1}$, \ie $\sim3$ times higher than either of the well studied, very radio luminous Type IIn SNe SN 1986J and SN 1988Z, and $\sim40$ times higher than the average peak 6 cm spectral luminosity of Type Ib/c SNe.  It also reached this 6 cm peak rather quickly,  only $\sim13$ days after explosion.

SN 1998bw is unusual in its radio emission, but not extreme.  For example, the time from explosion to peak 6 cm luminosity for both SN 1987A and SN 1983N was shorter  and, in spite of the fact that SN1998bw has been called ``the most luminous radio supernova ever observed,'' its  6 cm spectral luminosity  at peak is exceeded by that of SN 1982aa \cite{Yin94}.  However,  SN 1998bw is certainly the most radio luminous Type Ib/c radio supernova (RSN) observed so far by a factor of $\sim25$ and it reached this higher radio luminosity very early.

\subsection{Expansion Velocity \label{expansion}}

Although unique in neither the speed of radio light curve evolution nor in peak 6 cm radio luminosity, SN1998bw is certainly unusual in the combination of these two factors -- very radio luminous very soon after explosion.  Kulkarni \etal \cite{Kulkarni98-KW} have used these observed qualities, together with the lack of interstellar scintillation (ISS) at early times, brightness temperature estimates, and physical arguments to conclude that the blastwave from SN1998bw that gives rise to the radio emission must have been expanding relativistically.  On the other hand, Waxman \& Loeb \cite{Waxman99} argued that a sub-relativistic blastwave can generate the observed radio emission.  However, both sets of authors agree that a very high expansion velocity ($\gtrsim0.3c$) is required for the radio emitting region under a spherical geometry.

Simple arguments confirm this high velocity.  To avoid the well known Compton Catastrophe, Kellermann \& Pauliny-Toth \cite{Kellermann69} have shown that the brightness temperature $T_{\rm B} < 10^{12}$ K must hold and Readhead \cite{Readhead94} has better defined this limit to $T_{\rm B} < 10^{11.5}$ K.  From geometrical arguments, such a limit requires the radiosphere of SN1998bw to have expanded at an apparent speed $\gtrsim230,000$ \kms, at least during the first few days after explosion.  Although such a value is only mildly relativistic ($\Gamma \sim 1.6$; $\Gamma = ({1-\frac{v^2}{c^2}})^{-\frac{1}{2}}$), it is still unusually high.  However, measurements by Gaensler \etal \cite{Gaensler97-KW} and  Manchester \etal \cite{Manchester02} have demonstrated that the radio emitting regions of the Type II SN1987A expanded at an average speed of $\sim35,000$ \kms ($\sim0.1c$) over the 3 years from 1987 February to mid-1990 so that, in a very low density environment, such as one finds around Type Ib/c SNe, very high blastwave velocities appear to be possible.

\subsection{Radio Light Curves }

\begin{figure}[]
\rotatebox{-90}{\includegraphics[bb=2cm 1.25cm 20cm 19.25cm,width=9cm,height=9cm,clip=false]{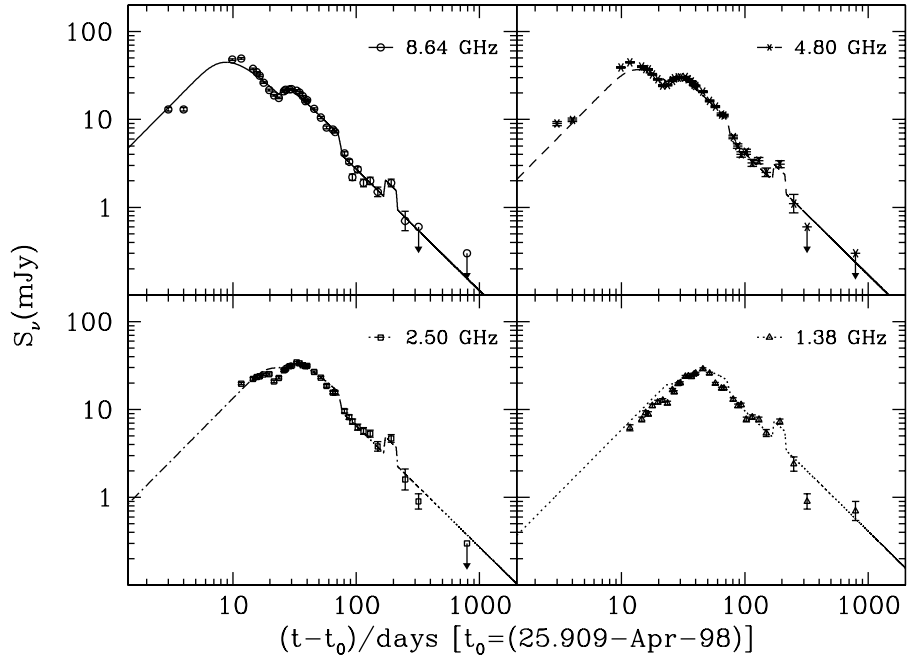}}
\caption[]{The radio light curves of SN 1998bw at 3.5 cm (8.6 GHz; upper left, {\it open circles, \it solid line}), 6.3 cm (4.8 GHz; upper right, {\it stars, dashed line}), 12 cm (2.5 GHz; lower left, {\it open squares, dash-dot line}) and 21 cm (1.4 GHz; lower right, {\it open triangles, dotted line}).  The curves are derived from a best fit model described by the Equations 1 -- 10 of \cite{Weiler01-KW} and the parameters and assumptions listed in Table \ref{tab1-KW}. During the 50 day intervals from day 25 - 75 and from day 165 - 215 the emission and absorption terms ($K_1$ and $K_3$; see \cite{Weiler01-KW}) increase by factors of 1.6 and 2.0, respectively, corresponding to a density increase of $40\%$ with a 6 day boxcar smoothed turn-on and turn-off of the enhanced emission/absorption.}
\label{fig1-KW}
\end{figure}

An obvious comparison of SN1998bw with other radio supernovae (RSNe) is the evolution of its radio flux density at multiple frequencies and its description by known RSN models (see \cite{Weiler01-KW}).  The radio data available at http://www.narrabri. atnf.csiro.au/$\sim$mwiering/grb/grb980425/ are plotted in Fig. \ref{fig1-KW}.  SN1998bw shows an early peak that reaches a maximum as early as day 10 -- 12 at 8.64 GHz, a minimum almost simultaneously for the higher frequencies ($\nu \ge$ 2.5 GHz) at day $\sim$ 20 -- 24, then a secondary, somewhat lower peak after the first dip.  An interesting characteristic of this ``double humped'' structure is that it dies out at lower frequencies and is relatively inconspicuous in the 1.38 GHz radio measurements (see Fig. \ref{fig1-KW}).

Such a ``double humped'' structure of the radio light curves can be reproduced by a single energy blastwave encountering differing CSM density regimes as it travels rapidly outward.  This is a reasonable assumption because complex density structure in the CSM surrounding SNe, giving rise to structure in the radio light curves, is very well known in such objects as SN1979C \cite{Montes00-KW,Weiler91-KW,Weiler92a-KW}, SN1980K \cite{Montes98-KW,Weiler92b-KW}, and, particularly, SN1987A \cite{Jakobsen91}.

Weiler \etal \cite{Weiler01-KW} pointed out what has not been previously recognized, that there is a sharp drop in the radio emission near day $\sim75$ and a single measurement epoch at day 192 that is significantly ($\sim60\%$) higher at all frequencies than expected from the preceding data on day 149 and the following data on day 249.

Weiler \etal ~were able to explain both of these temporary increases in radio emission by the SN blastwave encountering physically similar shells of enhanced density.  The first enhancement or ``bump'' after the initial outburst peak is estimated to start on day 25 and end on day 75, \ie having a duration of $\sim50$ days and turn-on and turn-off times of about 12 days, where the radio emission ($K_1$) increased by a factor of 1.6 and absorption ($K_3$) increased by a factor of 2.0 implying a density enhancement of $\sim40\%$ for no change in clump size.  Exactly the same density enhancement factor and length of enhancement is compatible with the ``bump'' observed in the radio emission at day 192 (\ie the single measurement within the 100 day gap between measurements on day 149 and day 249), even though the logarithmic time scale of Fig. \ref{fig1-KW} makes the time interval look much shorter.  The decreased sampling interval has only one set of measurements altered by the proposed day 192 enhancement, so Weiler \etal \cite{Weiler01-KW} could not determine its length more precisely than $<100$ days.

Li \& Chevalier \cite{Li99-KW} proposed an initially synchrotron self-absorbed, rapidly expanding blastwave in a $\rho \propto r^{-2}$ circumstellar wind model to describe the radio light curve for SN 1998bw.  This is in many ways similar to the Chevalier \cite{Chevalier98-KW} model for Type Ib/c SNe, that also included synchrotron self-absorption (SSA).  To produce the first bump in the radio light curves of SN1998bw, Li \& Chevalier postulated a boost of blastwave energy by a factor of $\sim2.8$ on day $\sim$22 in the observer's time frame.  They did not discuss the second bump.

Modeling of the radio data for SN 1998bw with the well established formalism for RSNe presented by Weiler \etal \cite{Weiler01-KW} shows that such an energy boost is not needed.  A fast blastwave interacting with a dense, slow, stellar wind-established, ionized CSM, that is modulated in density over time scales similar to those seen for RSNe, can produce a superior fit to the data.  No blastwave re-acceleration is required and no SSA at early times is apparent.  The parameters of the best fit model are given in Table~\ref{tab1-KW} and shown as the curves in Fig. \ref{fig1-KW}.  A visual comparison of the curves in Fig. \ref{fig1-KW} with those of Li \& Chevalier \cite{Li99-KW} Fig. 9, shows that the purely thermal absorption model with structured CSM provides a superior fit.

\begin{table}
\caption{SN1998bw/GRB980425 Modeling Results$^1$}\label{tab1-KW} 
\begin{center}
\begin{tabular}{lc} 
{Parameter}         & {Value} \\
$\alpha$ (spectral index)  & $-0.71$ \\ 
$\beta$ (decline rate)     & $-1.38$ \\ 
$K_1^a$               & $2.4 \times 10^3$ \\  
$K_2$               & 0  \\
$\delta$            & -- \\
$K_3^a$               & $1.7 \times 10^3$ \\
${\delta}^{\prime}$ & $-2.80$ \\
$K_4$               & $1.2 \times 10^{-2}$ \\
$t_0$(Explosion Date) & 1998 Apr. 25.90915 \\
($t_{\rm 6cm\ peak} - t_0$)(days) & 13.3 \\
$S_{\rm 6cm\ peak}$(mJy) & 37.4 \\
d(Mpc) & 38.9 \\
$L_{\rm 6cm~peak}\ ({\rm ergs\ s^{-1}\ Hz^{-1}})$ & $6.7 \times 10^{28}$ \\
${\dot M}({\rm M_\odot} ~ {\rm yr}^{-1})^b$ & $2.6 \times 10^{-5}$ \\
\end{tabular}
\end{center}
$^1$See Weiler \etal \cite{Weiler01-KW} for parameter definitions.\\
$^a$Enhanced by a factors of $1.6$ ($K_1$) and 2.0 ($K_3$), corresponding to a density increase of $40\%$, over the intervals day 25 - 75 and day 165 - 215, although the latter interval could be as long as 100 days and still be compatible with the available data.\\
$^b$Assuming $t_{\rm i} = 23~{\rm days}$, $t = (t_{\rm 6cm\ peak} - t_0) = 13.3$ days, $m = -(\alpha - \beta - 3)/3 = 0.78$, $w_{\rm wind}$ = 10 \kms, $v_{\rm i} = v_{\rm blastwave} = 230,000$ \kms, $T = 20,000$ K, average number of clumps along the line-of-sight $N = 0.5$ and volume filling factor $\phi = 0.22$ (see Weiler \etal \cite{Weiler01-KW}, Case 3).
\end{table}

One should note that the fit listed in Table~\ref{tab1-KW} and shown as the curves in Fig. \ref{fig1-KW} requires no ``uniform'' absorption ($K_2 = 0$) so all of the free-free (f-f) absorption is due to a clumpy medium as described in Equations 1 and 5 of \cite{Weiler01-KW}.  These results, combined with the estimate of a high blastwave velocity, suggest that the CSM around SN1998bw is highly structured with little, if any, inter-clump gas.  The clump filling factor must be high enough to intercept a considerable fraction of the blastwave energy and low enough to let radiation escape from any given clump without being appreciably absorbed by any other clump which is Case 3 discussed by Weiler \etal \cite{Weiler01-KW}. The blastwave can then easily move at a speed that is a significant fraction of the speed of light, because it is moving in a very low density medium, but still cause strong energy dissipation and relativistic electron acceleration at the clump surfaces facing the SN explosion center. 

Weiler \etal \cite{Weiler01-KW} also noted from the fit given in Table~\ref{tab1-KW} that the presence of a $K_4$ (see \cite{Weiler01-KW}) factor implies there is a more distant, uniform screen of ionized gas surrounding the exploding system that is too far to be affected by the rapidly expanding blastwave and provides a time independent absorption.

\subsection{Physical Parameter Estimates}

Using the fitting parameters from Table~\ref{tab1-KW} and Equations 11 and 16 of \cite{Weiler01-KW}, Weiler \etal \cite{Weiler01-KW} estimated a mass-loss rate from the preexplosion star.  The proper parameter assumptions are rather uncertain for these enigmatic objects but, for a preliminary estimate, they assumed $t_{\rm i} = 23~{\rm days}$, $t = (t_{\rm 6cm\ peak} - t_0) = 13.3$ days, $m = -(\alpha - \beta - 3)/3 = 0.78$ (Equation 7 of \cite{Weiler01-KW}), $w_{\rm wind}$ = 10 \kms (for an assumed RSG progenitor), $v_{\rm i} = v_{\rm blastwave} = 230,000$ \kms, and $T = 20,000$ K.  They also assumed, because the radio emission implies that the CSM is highly clumped (ie $K_2 = 0$), that the CSM volume is only sparsely occupied ($N = 0.5$, $\phi = 0.22$; see Weiler \etal \cite{Weiler01-KW}, Case 3).  Within these rather uncertain assumptions, Equations 11 and 16 of Weiler \etal \cite{Weiler01-KW} yield an estimated mass-loss rate of ${\dot M} \sim2.6 \times 10^{-5}\ {\rm M_\odot}\ {\rm yr}^{-1}$ with density enhancements of $\sim40\%$  during the two known, extended ``bump'' periods.

Assuming that the blastwave is traveling at a constant speed of $\sim230,000$ \kms the beginning of the first ``bump'' on day 25 implies that it starts at $\sim5.0 \times 10^{16}$ cm and ends on day 75 at $\sim1.5 \times 10^{17}$ cm from the star.  Correspondingly, if it was established by a 10 \kms RSG wind, the 50 days of enhanced mass-loss ended $\sim1,600$\ yr and started $\sim4,700$\ yr before the explosion.  The earlier high mass-loss rate epoch indicated by the enhanced emission on day 192 in the measurement gap between day 149 and day 249 implies, with the same assumptions, that it occurred in the interval between $\sim9,400$ yr and $\sim15,700$ yr before explosion.  It is interesting to note that the time between the presumed centers of the first and second increased mass-loss episodes of $\sim9,400$\ yr is comparable to the $\sim12,000$\ yr before explosion at which SN1979C had a significant mass-loss rate increase \cite{Montes00-KW} and SN1980K had a significant mass-loss rate decrease \cite{Montes98-KW}, thus establishing a possible characteristic time scale of $\sim10^4$\ yr for significant changes in mass-loss rate for preexplosion massive stars.  

\subsection{Radio Emission from SN1998bw/GRB980425 and Other GRB Radio Afterglows}

Because the suggestion of a possible relation between SN1998bw and GRB 980425, it has remained a tantalizing possibility that the origin of at least some GRBs is in the better known Type Ib/c SN phenomenon.  First, of course, one must keep in mind that there may be (and probably is) more than one origin for GRBs, a situation that is true for most other classes of objects.  For example, SNe, after having been identified as a new phenomenon in the early part of the last century, were quickly split into several subgroups such as Zwicky's Types I -- V, then coalesced back into just two subgroupings based on H$\alpha$ absent (Type I) or H$\alpha$ present (Type II) in their optical spectra.  This simplification has not withstood the test of time, however, and subgroupings of Type Ia, Ib, Ic, II, IIb, IIpec, IIn, and others have come into use over the past 20 years.

GRBs, although at a much earlier stage of understanding, have similarly started to split into subgroupings.  The two currently accepted groupings are referred to as ``fast-hard'' and ``slow-soft'' from the tendency of the $\gamma$-ray emission for some to evolve more rapidly (mean duration $\sim$0.2 s) and to have a somewhat harder spectrum than for others that evolve more slowly  (mean duration $\sim$20 s) with a somewhat softer spectrum \cite{Fishman95-KW}.

Because we are only concerned with the radio afterglows of GRBs here, all of our examples fall into the slow-soft classification, at least partly because the fast-hard GRBs fade too quickly for followup observations to obtain the precise positional information needed for identification at longer wavelengths.  It is therefore uncertain whether fast-hard GRBs have radio afterglows or even whether the slow-soft GRBs represent a single phenomenon.  If, however, we assume that at least some types of slow-soft GRBs have a similar origin and that GRB980425/SN1998bw is a key to this puzzle telling us that such ``slow-soft'' GRBs have their origin in at least some types of SNe, we can investigate relations between the two observational phenomena.

\section{Gamma-Ray Bursters}

\subsection{Radio Detections \label{detections}}

Beyond GRB980425, there are relatively few GRBs with detected radio afterglows and only six of these as of 2001 Dec. 31 have sufficient radio light curve information to permit approximate model fits.  Additionally, because the optical/infrared (OIR) data appear consistent with a synchrotron origin similar to that of the radio emission, we have collected the available OIR data and included it in our  model fitting to better constrain the source parameters.  Although detailed modeling is beyond the scope of this review, we apply the parameterization of Equations 1 to 10 of Weiler \etal \cite{Weiler01-KW} to the available radio and OIR data in attempt to highlight some of the gross properties of the GRB afterglow processes.  Because the OIR data suffer an extinction that is absent in the radio, a zero redshift color excess $[E(B-V)]$ was also obtained from the fitting by adopting the Galactic law of Savage \& Mathis \cite{Savage79}. 

{\bf GRB970508} was discovered by the \B team on 1997 May 8.904 UT \cite{Costa97b}.  These results showed detection of an afterglow in all wavelength bands including X-ray \cite{Piro97}, optical \cite{Bond97}, and radio \cite{Frail97a}.  Frail \& Kulkarni \cite{Frail97a} derived a position from their 8.46 GHz VLA observations of RA(J2000) = $06^h 53^m 49\fs45$, Dec(J2000) = $+79\arcdeg 16\arcmin 19\farcs5$ with an error of $\sim$0\farcs1 in both coordinates.  Metzger \etal \cite{Metzger97-KW} found a redshift of $z = 0.835$, which Bloom \etal \cite{Bloom98} confirmed for the host galaxy. 

\begin{figure}[]
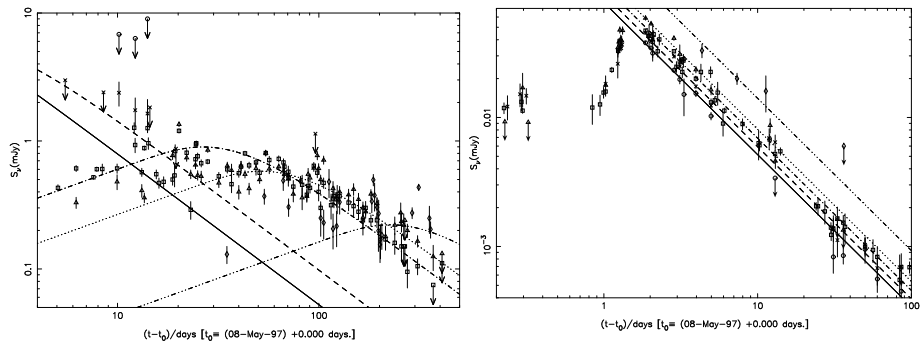

\begin{center}
\includegraphics[angle=-90,width=6.0cm]{Weilerfig2a.ps}
\includegraphics[angle=-90,width=6.0cm]{Weilerfig2b.ps}
\end{center}
\caption[]{GRB970508 at radio wavelengths (left, 2a) of 1.3 mm (232 GHz; {\it open circles, solid line}), 3.5 mm (86.7 GHz; {\it crosses, dashed line}), 3.5 cm (8.5 GHz; {\it open squares, dash-dot line}), 6 cm (4.9 GHz; {\it open triangles, dotted line}), and 20 cm (1.5 GHz; {\it open diamonds, dash-triple dot line}) and optical/IR bands (right, 2b) of B (681 THz; {\it circles, solid line}), V (545 THz; {\it crosses, dashed line}), R (428 THz; {\it open squares, dash-dot line}), I (333 THz; {\it open triangles, dotted line}), and K (138 THz; {\it open diamonds, dash-triple dot line}).  (NB: To enhance clarity, the scales are different between the two figures even though the fitting parameters are the same and not all available bands are plotted even though they were used in the fitting.)}
\label{fig2-KW}
\end{figure}

The radio data were obtained from the references \cite{Bremer98,Frail00b,Galama98c,Shepherd98,Smith99} and the OIR data were obtained from the references \cite{Castro98,Chary98,Djorgovski97,Galama98c,Galama98d,Garcia98,Sahu97-KW,Schaefer97,Sokolov98}.  Representative data for GRB970508 are plotted, along with curves from the best fit model, in Fig. \ref{fig2-KW} and the parameters of the fit are listed in Table~\ref{tab2-KW}.

\begin{table}
\caption{Fitting Parameters for GRB Radio Afterglows$^1$}\label{tab2-KW}
\begin{center}
\scriptsize
\renewcommand{\arraystretch}{0.8}
\begin{tabular}{lccccccccc} 
GRB & \multicolumn{2}{c}{$E(B-V)$} & Redshift$^4$ & Spectral &  Decline & $K_1^5$ & $K_3^5$ & $\delta^\prime$ & Peak 6 cm \\
 & Milky   & Host       & & Index & Rate & & & & Radio      \\
 & Way$^2$ & Galaxy$^3$ & & & & & & & Luminosity$^5$ \\
 & (mag)& (mag) & (z) &($\alpha$) &($\beta$) & & & & (${\rm erg~s^{-1} Hz^{-1}}$) \\
\hline
GRB970508& 0.08 & 0.00 & 0.835& $-0.63$ & $-1.18$ & $1.35 \times 10^{2}$ & $1.81 \times 10^{3}$ & $-1.75$ & $1.39 \times 10^{31}$ \\
GRB980329 & 0.07 & 0.28 & $\equiv1.000$ & $-1.33$ & $-1.09$ & $1.54 \times 10^{4}$ & $1.33 \times 10^{5}$ & $-1.16$ & $7.69 \times 10^{30}$ \\
GRB980425$^a$& -- & -- & 0.0085 & $-0.71$ & $-1.38$ & $2.37 \times 10^{3}$ & $1.73 \times 10^{3}$ & $-2.80$ & $6.70 \times 10^{28}$ \\
GRB980519 & 0.27 & 0.00 & $\equiv1.000$ & $-0.75$ & $-2.08$ & $8.45 \times 10^{1}$ & $1.37 \times 10^{4}$ & $-3.57$ & $7.52 \times 10^{30}$ \\
GRB991208 & 0.02 & 0.07 & 0.707 & $-0.58$ & $-2.27$ & $5.11 \times 10^{2}$ & $3.53 \times 10^{3}$ & $-3.29$ & $2.07 \times 10^{31}$ \\
GRB991216 & 0.63 & 0.11 & 1.020 & $-0.28$ & $-1.38$ & $7.07 \times 10^{0}$ & $2.50 \times 10^{1}$ & $-1.40$ & $1.05 \times 10^{31}$ \\
GRB000301C$^b$& 0.05 & 0.05 & 2.034 & $-0.60$ & $\equiv-1.75$ & $2.31 \times 10^{2}$ & $2.72 \times 10^{3}$ & $-2.06$ & $3.77 \times 10^{31}$ 
 \\
\end{tabular}
\end{center}
$^1$The fits do not generally require use of $K_2, \delta, K_4, K_5, \delta^{\prime \prime}, K_6, \delta^{\prime \prime \prime}$ (see Weiler \etal \cite{Weiler01-KW}) so that these parameters are not included.\\
$^2$Galactic extinction in the direction of the $\gamma$-ray burster (GRB) was obtained from \cite{Halpern00-KW,Reichart99,Schlegel98}.\\
$^3$Additional extinction, at the host galaxy redshift (see $\S$\ref{detections}), is needed in some cases to provide a good fit to the optical/infrared (OIR) data.\\
$^4$Where unknown, the redshift is defined to be $z = 1.000$.\\
$^5$Derived for 6 cm in the rest frame of the observer, not the emitter.\\
$^a$The best fit includes a $K_4 = 1.24 \times 10^{-2}$ term.\\
$^b$The OIR data and radio data appear to have different rates of decline ($\beta_{\rm OIR} \sim -2.0$, $\beta_{\rm radio} \sim -1.5$) implying that there may be a break between the two regimes.  For the purposes of this review the average of $\beta  \equiv -1.75$ has been adopted.
\end{table}

Examination of Figs. \ref{fig2-KW}a,b shows that the parameterization listed in Table~\ref{tab2-KW} describes the data well in spite of the very large frequency and time range.  In Fig. \ref{fig2-KW}a the 232 GHz upper limits and the 86.7 GHz limits and detections are in rough agreement with the model fitting; the 8.5 GHz and 4.9 GHz measurements are described very well, and even the 1.5 GHz data are consistent with the parameterization if significant interstellar scintillation (ISS) is present (see $\S$ \ref{scintillation}).  Waxman \etal \cite{Waxman98} have already ascribed the large fluctuations in the flux density at both 8.5 and 4.9 GHz to ISS, and one expects such ISS to also be present at 1.4 GHz.  

In Fig. \ref{fig2-KW}b, although the OIR data show significant scatter at individual frequencies, the data are consistent with a non-thermal, synchrotron origin that has the same decline rate ($\beta$) and spectral index ($\alpha$) as in the radio regime, indicating that no spectral breaks have occurred between the two observing ranges.  The color excess of $E(B-V) = 0.08$ mag from Schlegel \etal \cite{Schlegel98} is consistent with the best fit to the data, implying little extinction in the host galaxy.

The most obvious characteristic of the OIR data is that in the initial interval between the time of the GRB and day 1.75 the data are not well modeled by the parameterization.  This is not surprising, because the modeling contains no parameters to describe any turn-on effects for the synchrotron emission.  Thus, the initial OIR data prior to day 1.75 are particularly suitable for constraining explosion models such as those discussed in $\S$ \ref{models}.

{\bf GRB980329} was discovered by the \B team (BATSE Trigger \# 6665, \cite{Briggs98}) on 1998 Mar.~29.1559 UT \cite{Frontera98}.  The afterglow was detected in all wavelength bands including X-ray \cite{Zand98-KW}, optical \cite{Klose98}, and radio \cite{Taylor98a}.  Taylor \etal \cite{Taylor98a} derived a position from their VLA observations of RA(J2000) = $07^h 02^m 38\fs022$, Dec(J2000) = $+38\arcdeg 50\arcmin 44\farcs02$ with an uncertainty of $\pm 0\farcs05$ in each coordinate.  Unfortunately, no redshift has been obtained for the optical afterglow of GRB980329 or its inferred parent galaxy so that, except for arguments that is quite distant with, perhaps, z $\sim$ 5, no reliable distance estimate is available.

\begin{figure}[t]
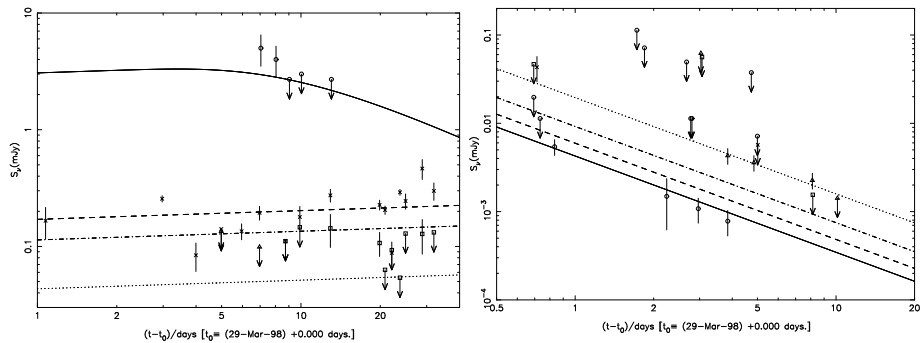

\begin{center}
\includegraphics[angle=-90,width=6.0cm]{Weilerfig3a.ps}
\includegraphics[angle=-90,width=6.0cm]{Weilerfig3b.ps}
\end{center}
\caption[]{GRB980329 at radio wavelengths (left, 3a) of 0.9 mm (352 GHz; {\it open circles, solid line}), 3.5 cm (8.5 GHz; {\it crosses, dashed line}), 6 cm (4.9 GHz; {\it open squares, dash-dot line}), and 20 cm (1.4 GHz; {\it open triangles, dotted line}) and optical/IR bands (right, 3b) of R (428 THz; {\it open circles, solid line}), I (333 THz; {\it crosses, dashed line}), J (240 THz; {\it open squares, dash-dot line}), and K (136 THz; {\it open triangles, dotted line}).  (NB: To enhance clarity, the scales are different between the two figures even though the fitting parameters are the same and not all available bands are plotted even though they were used in the fitting.)}
\label{fig3-KW}
\end{figure}

The radio data were obtained from \cite{Smith98,Smith99,Taylor98b} and the OIR data were obtained from \cite{Gorosabel99-KW,Reichart99}.  Representative data for GRB980329 are plotted, along with curves from the best fit model, in Fig. \ref{fig3-KW} and the parameters of the fit are listed in Table~\ref{tab2-KW}.

Examination of Figs. \ref{fig3-KW}a,b shows that the parameterization listed in Table~\ref{tab2-KW} describes the data rather well over the broad parameter space in time and frequency.  In Fig. \ref{fig3-KW}a the 352 GHz detections and upper limits are in good agreement with the parameterization; the 8.5 GHz and 4.9 GHz measurements are described very well although there may be some ISS present (see $\S$ \ref{scintillation}); the 1.4 GHz upper limits are consistent with the parameterization.

In Fig. \ref{fig3-KW}b the data are mostly upper limits, with only a few detections at K-band (136 THz) and R-band (428 THz).  However, these are surprisingly well described by the parameterization if a zero redshift color excess of $E(B-V) = 0.75$ mag is assumed.  This value for color excess is much higher than the value of $E(B-V) = 0.073$ mag from Reichart \etal \cite{Reichart99} for Galactic color excess in that direction.  However, if we assume that the additional color excess arises in the host galaxy and that the color excess scales with redshift as $(1+z)^{-1.25}$, as is appropriate for an adopted Small Magellanic Cloud extinction law \cite{Prevot84}, then the host galaxy contributes $E(B-V) = 0.28$ mag (Table~\ref{tab2-KW}).  Such a color excess is fairly typical for the disk of a late-type galaxy.

Taylor \etal \cite{Taylor98b} have also modeled the radio data and invoked a somewhat steeper, inverted spectrum with $\alpha = -1.7$ between 4.9 and 8.3 GHz, flattening to $\alpha = -0.8$ between 15 and 90 GHz caused by a SSA component with a turnover frequency near 13 GHz.  Extrapolating to higher frequencies, their model predicts a rather low 350 GHz flux density of only $\sim$1.7 mJy that is incompatible with the \J measurements.

Such complexity is not be needed, however.  Our parameterization listed in Table \ref{tab2-KW} and shown in Fig. \ref{fig3-KW}a yields a good description of the data and predicts a 350 GHz flux density of $\sim$3.0 mJy, in much better agreement with the observations. 

Our model fit to the available first 30 days of radio data indicates that GRB980329 should be detectable with the VLA at centimeter wavelengths, with little decline, for an extended period.  Although prediction of exact flux densities at later times is not reliable because the decline phase of the model is not well constrained by the few available data, it is interesting to note that GRB980329 has apparently been detected by the VLA at 8.46 GHz, 4.86 GHz, and 1.43 GHz for up to 500 days after the outburst \cite{Young99}.  

{\bf GRB980519} was discovered by the \B team (BATSE Trigger \# 6764) on 1998 May 19.51410 UT \cite{Muller98}.  The afterglow was detected in all wavelength bands including X-ray \cite{Nicastro98}, optical \cite{Jaunsen98}, and radio \cite{Frail98}.  Frail \etal \cite{Frail98} derived a position from their VLA observations of  RA(J2000) = $23^h 22^m 21\fs49$, Dec(J2000) = $+77\arcdeg 15\arcmin 43\farcs2$ with an uncertainty of $\pm 0\farcs1$ in each coordinate.  Unfortunately, no redshift has been obtained for the optical afterglow so that no distance estimate is available.

\begin{figure}[t]
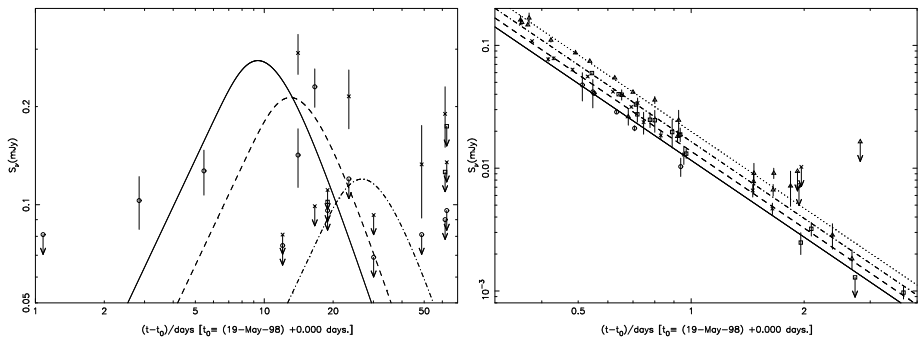

\begin{center}
\includegraphics[angle=-90,width=6.0cm]{Weilerfig4a.ps}
\includegraphics[angle=-90,width=6.0cm]{Weilerfig4b.ps}
\end{center}
\caption[]{GRB980519 at radio wavelengths (left, 4a) of 3.5 cm (8.5 GHz; {\it open circles, solid line}), 6 cm (4.9 GHz; {\it crosses, dashed line}), and 20 cm (1.4 GHz; {\it open squares, dash-dot line}) and optical/IR bands (right, 4b) of B (681 THz; {\it open circles, solid line}), V (545 THz; {\it crosses, dashed line}), R (428 THz; {\it open squares, dash-dot line}), and I (333 THz; {\it open triangles, dotted line}).  (NB: To enhance clarity, the scales are different between the two figures even though the fitting parameters are the same and not all available bands are plotted even though they were used in the fitting.)}
\label{fig4-KW}
\end{figure}

The radio data were obtained from \cite{Frail00b} and the OIR data from \cite{Halpern99,Jaunsen01,Vrba00}  Representative data for GRB980519 are plotted, along with curves from the best fit model, in Fig. \ref{fig4-KW} and the parameters of the fit are listed in Table~\ref{tab2-KW}. 

Examination of Figs. \ref{fig4-KW}a,b shows that the parameterization listed in Table~\ref{tab2-KW} describes the data reasonably well, even though the radio data shown in Fig. \ref{fig4-KW}a are very limited and of relatively poor quality.  For example, the reported detections at 4.9 GHz after day 50 are barely more than 3$\sigma$ and thus of limited reliability, so that the fit is not well constrained and the significance of their deviation from the best fit curve is unknown.  The data at both 8.5 GHz and 4.9 GHz have significant fluctuations yielding detections and 3$\sigma$ upper limits at relatively small time separations so that GRB980519 may be undergoing ISS (see $\S$ \ref{scintillation}).  Only upper limits are available at 1.4 GHz, but they are consistent with the best fit model.

In Fig. \ref{fig4-KW}b there is good coverage in both frequency and time. Although there is some fluctuation, the data are rather well described by the parameterization if a color excess of $E(B-V) = 0.25$ mag is assumed.  This is rather close to the value of $E(B-V) = 0.267$ mag suggested by Schlegel \etal \cite{Schlegel98} in that direction from their \I 100 $\mu$m maps.

{\bf GRB991208} was discovered by the Ulysses and KONUS and NEAR teams on 1999 Dec.~08.1923 UT \cite{Hurley99}.  The afterglow was detected in the optical \cite{Castro99} and radio \cite{Frail99b} wavelength bands.  There does not appear to have been an X-ray detection.  Frail \& Kulkarni \cite{Frail99b} derived an position for the radio transient of RA(J2000) = $16^h 33^m 53\farcs50$, Dec(J2000) = $+46\arcdeg 27\arcmin 20\farcs9$ with no error given, but normally for the VLA positional errors are $<0\farcs1$.  Dodonov \etal \cite{Dodonov99} found a redshift for the parent galaxy of $z = 0.707 \pm 0.002$. 

\begin{figure}[t]
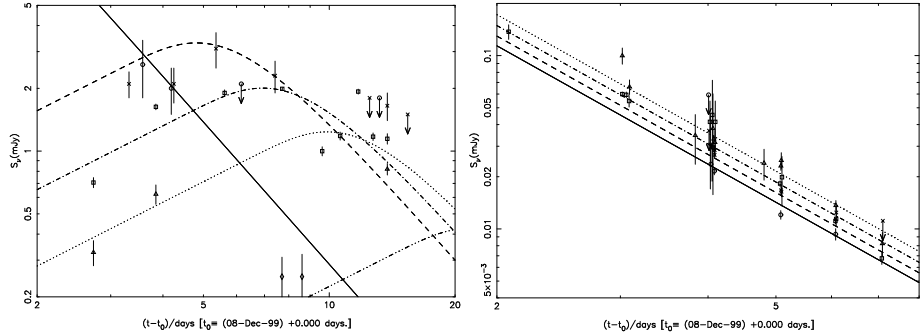

\begin{center}
\includegraphics[angle=-90,width=6.0cm]{Weilerfig5a.ps}
\includegraphics[angle=-90,width=6.0cm]{Weilerfig5b.ps}
\end{center}
\caption[]{GRB991208 at radio wavelengths (left, 5a) of 1.2 mm (250 GHz; {\it open circles, solid line}), 2.0 cm (15.0 GHz; {\it crosses, dashed line}), 3.5 cm (8.5 GHz; {\it open squares, dash-dot line}),  6 cm (4.9 GHz; {\it open triangles, dotted line}), and 20 cm (1.4 GHz; {\it open diamonds, dash-triple dot line}) and optical/IR bands (right, 5b) of B (681 THz; {\it open circles, solid line}), V (545 THz; {\it crosses, dashed line}), R (428 THz; {\it open squares, dash-dot line}), and I (333 THz; {\it open triangles, dotted line}).  (NB: To enhance clarity, the scales are different between the two figures even though the fitting parameters are the same and not all available bands are plotted even though they were used in the fitting.)}
\label{fig5-KW}
\end{figure}

The radio data were obtained from Galama \etal \cite{Galama00} and the OIR data from Castro-Tirado \etal \cite{Castro01}.  Representative data for GRB991208 are plotted, along with curves from the best fit model, in Fig. \ref{fig5-KW} and the parameters of the fit are listed in Table~\ref{tab2-KW}.

Examination of Figs. \ref{fig5-KW}a,b shows that the parameterization listed in Table~\ref{tab2-KW} describes the data reasonably well.  In Fig. \ref{fig5-KW}a the radio data are quite limited and have relatively large scatter.  For example, the reported detection at 15 GHz on day 7.4 is preceded by a $3\sigma$ upper limit of comparable magnitude on day 5.4 and followed by an upper limit on day 12.5, so that the reality of the day 7.4 detection must be called into question.  In any case, although the fit is not well constrained by the radio data, the model curves describe its evolution reasonably well.  The data are too sparse to judge if ISS is present.

In Fig. \ref{fig5-KW}b there is good coverage in both frequency and time. There is some fluctuation, particularly for the four R-band observations on day 4.  However, examination of the data reveals that the three high R-band measurements on day 4.026, 4.058, and 4.096 have significantly larger errors ($\sim0.4$ mag) than the R-band datum on day 4.068 (error = 0.050 mag) that is consistent with the fitted curves.  In general, the data are rather well described by the parameterization if a zero redshift color excess of $E(B-V) = 0.15$ mag is assumed.  This is appreciably higher than the value of $E(B-V) = 0.016$ mag suggested by Schlegel \etal \cite{Schlegel98} in that direction from their \I 100 $\mu$ maps, implying, under the same assumptions as were used for GRB980329 above, a $E(B-V) = 0.07$ mag at the redshift of the host galaxy (Table~\ref{tab2-KW}).

{\bf GRB991216} was discovered by BATSE on 1999 Dec.~16.671544 UT (BATSE trigger \# 7906) \cite{Kippen99a}.  The afterglow was detected in all wavelength bands including x-ray \cite{Takeshima-etal-99}, optical \cite{Uglesich99}, and radio \cite{Taylor99a}.  Taylor \& Berger \cite{Taylor99a} derived a position from their 8.5 GHz VLA observations of RA(J2000) = $05^h 09^m 31\fs297$, Dec(J2000) = $+11\arcdeg 17\arcmin 07\farcs25$ with an error of $\pm 0\farcs1$ in each coordinate. Vreeswijk \etal \cite{Vreeswijk99} suggested a redshift of $z \ge 1.02$ based on the highest redshift of three possible absorption systems seen in the optical afterglow.

\begin{figure}[t]
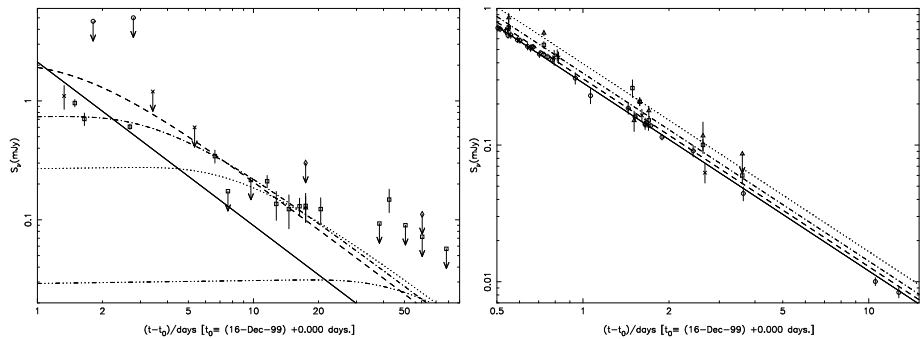

\begin{center}
\includegraphics[angle=-90,width=6.0cm]{Weilerfig6a.ps}
\includegraphics[angle=-90,width=6.0cm]{Weilerfig6b.ps}
\end{center}
\caption[]{GRB991216 at radio wavelengths (left, 6a) of 0.9 mm (350 GHz; {\it open circles, solid line}), 2.0 cm (15.0 GHz; {\it crosses, dashed line}), 3.5 cm (8.5 GHz; {\it open squares, dash-dot line}), 6 cm (4.9 GHz; {\it open triangles, dotted line}),and 20 cm (1.4 GHz; {\it open diamonds, dash-triple dot line}), and optical/IR bands (right, 6b) of R (428 THz; {\it open circles, solid line}), I (333 THz; {\it crosses, dashed line}), J (239 THz; {\it open squares, dash-dot line}), and K (136 THz; {\it open triangles, dotted line}).  (NB: To enhance clarity, the scales are different between the two figures even though the fitting parameters are the same and not all available bands are plotted even though they were used in the fitting.)}
\label{fig6-KW}
\end{figure}

The radio data were obtained from Frail \etal \cite{Frail00c} and the OIR data from Halpern \etal \cite{Halpern00-KW}, Frail \etal \cite{Frail00c}, and Garnavich \etal \cite{Garnavich00}. Representative data for GRB991216 are plotted, along with curves from the best fit model, in Fig. \ref{fig6-KW} and the parameters of the fit are listed in Table~\ref{tab2-KW}.

Examination of Figures \ref{fig6-KW}a,b shows that the parameterization listed in Table~\ref{tab2-KW} fits the data reasonably well, even with the radio data (Fig. \ref{fig6-KW}a) being quite limited and with relatively large scatter.  The single detection at 8.5 GHz on day 42.49 surrounded by much lower $3\sigma$ upper limits on days 38.28 and 50.51, if correct, is not well described.  The data are too sparse to judge if ISS is observed.

In Fig. \ref{fig6-KW}b there is reasonably good coverage in both frequency and time, particularly at R-band. A satisfactory fit to the data requires a rather high zero redshift color excess of $E(B-V) = 0.90$ mag.  This is significantly higher than the value of $E(B-V) = 0.63$ mag suggested by Schlegel \etal \cite{Schlegel98} in that direction from their \I 100 $\mu$ maps or the value of $E(B-V) = 0.40$ mag  derived by Halpern \etal \cite{Halpern00-KW} from the Galactic 21 cm column density in that direction.  If, as above, we assume that the additional extinction arises in the host galaxy, then with the same assumptions as were used for GRB980329, we obtain a color excess of $E(B-V) = 0.11$ mag at the redshift of the host galaxy (Table~\ref{tab2-KW}).

{\bf GRB000301C} was discovered by the Ulysses and NEAR teams on 2000 Mar.~ 01.4108 UT  \cite{Smith00}.  The afterglow was detected in the optical \cite{Fynbo00} and radio \cite{Bertoldi00} wavelength bands.  There does not appear to have been an X-ray detection.  Fynbo \etal \cite{Fynbo00} derived a position for the optical transient of RA(J2000) = $16^h 20^m 18\farcs6$, Dec(J2000) = $+29\arcdeg 26\arcmin 36\arcsec$ with an error of $\sim1\arcsec$ in both coordinates.  Castro \etal \cite{Castro00} (see also \cite{Smette00}) found a redshift for the parent galaxy of $z = 2.0335 \pm 0.003$.

\begin{figure}[t]
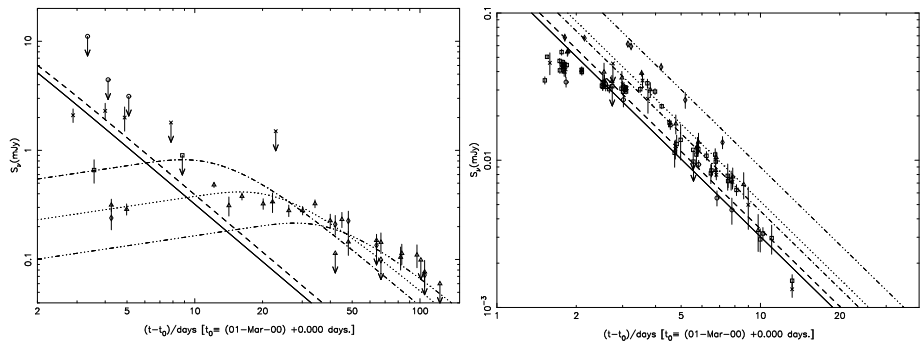

\begin{center}
\includegraphics[angle=-90,width=6.0cm]{Weilerfig7a.ps}
\includegraphics[angle=-90,width=6.0cm]{Weilerfig7b.ps}
\end{center}
\caption[]{000301C at radio wavelengths (left, 7a) of 0.9 mm (350 GHz; {\it open circles, solid line}), 1.2 mm (250 GHz; {\it crosses, dashed line}), 2.0 cm (15.0 GHz; {\it open squares, dash-dot line}),  3.5 cm (8.5 GHz; {\it open triangles, dotted line}), and 6.0 cm (4.9 GHz; {\it open diamonds, dash-triple dot line}) and optical/IR bands (right, 7b) of U (832 THz; {\it circles, solid line}), B (681 THz; {\it crosses, dashed line}), R (428 THz; {\it open squares, dash-dot line}), I (333 THz; {\it open triangles, dotted line}), and K (136 THz; {\it open diamonds, dash-triple dot line}).  (NB: To enhance clarity, the scales are different between the two figures even though the fitting parameters are the same and not all available bands are plotted even though they were used in the fitting.)}
\label{fig7-KW}
\end{figure}

The radio data were obtained from Berger \etal \cite{Berger00} and the OIR data from Sagar \etal \cite{Sagar00}, Bhargavi \& Cowsik \cite{Bhargavi00}, Masetti \etal \cite{Masetti00}, Rhoads \& Fruchter \cite{Rhoads01}, and Jensen \etal \cite{Jensen01}.  Representative data for GRB000301C are plotted, along with curves from the best fit model, in Fig. \ref{fig7-KW} and the parameters of the fit are listed in Table~\ref{tab2-KW}.

Examination of Figures ~\ref{fig7-KW}a,b shows that the parameterization listed in Table~\ref{tab2-KW} is successful in that it describes the data rather well over the large frequency and time range.  In Fig. \ref{fig7-KW}a the 350 GHz upper limits and the 250 GHz detections and limits are in rough agreement with the parameterization; the 15 GHz and 8.5 GHz measurements are also consistent with the modeling, although the model drops off a bit faster at late times than the data and the fit would be improved by a somewhat flatter decline rate of $\beta \sim -1.5$.  The 1.4 GHz data are generally consistent with the model, although ISS (see $\S$ \ref{scintillation}) may be present in the first measurement.

In Fig. \ref{fig7-KW}b, although the OIR data show significant scatter at individual frequencies, they are consistent with a non-thermal, synchrotron origin.  A decline rate of $\beta \sim -2.0$, somewhat steeper than in the radio, would provide an improved fit.  The fit indicates a best value for the zero redshift color excess of $E(B-V) = 0.25$ mag, significantly higher than the value of $E(B-V) = 0.052$ mag suggested by Schlegel \etal \cite{Schlegel98} in that direction from their \I 100 $\mu$ maps.  As above, this implies a color excess of $E(B-V) = 0.05$ mag at the redshift of the host galaxy (Table~\ref{tab2-KW}).

The most obvious characteristic of the OIR data is that the initial interval between the time of the GRB and day 3.5, is not well modeled by the parameterization.  This is not surprising, because the modeling contains no parameters to describe any turn-on effects for the synchrotron emission.  A similar turn-on effect was seen for the early OIR data for GRB970508, although not for the other GRBs discussed here.  

It should be noted that the possible steepening of the flux density decline rate from $\beta \sim -1.5$ (radio) to $\beta \sim -2.0$ (OIR) may indicate a break in the decline rate somewhere between the two observing ranges, possibly owing to synchrotron losses of the high energy electrons.

\subsection{Parameterization}

Even though the study of the GRB phenomenon is still very much in its early stages and, as has been pointed out above, the GRBs for which we have any afterglow information are all of the ``slow-soft'' category with those about which we have significant radio information only a small subset, the parameterization studies (see Table \ref{tab2-KW}) allow us to draw a some tentative conclusions:

\begin{enumerate}
\item{Of the seven relatively well observed radio GRBs (RGRBs)(including GRB 980425/SN1998bw), five appear to have similar relativistic electron acceleration processes that generate a spectral index $\alpha \sim -0.6$ to $\sim -0.7$.  The two possible exceptions are GRB980329 ($\alpha \sim -1.3$) and GRB991216 ($\alpha \sim -0.3$) which, interestingly enough, straddle the average spectral index of the other objects.   The average spectral index value is very similar to the canonical value for extragalactic radio sources.}
\item{The RGRBs can be described rather well from radio through OIR by the same formalism of Equations 1 to 10 of Weiler \etal \cite{Weiler01-KW} that is successful in describing the radio emission from RSNe as arising from a blastwave/CSM interaction.  The obvious conclusion is that at least this subset of GRBs arise from the explosion of a compact, presumably stellar-sized object embedded in a dense CSM.}
\item{Only two of the objects, GRB970508 and GRB000301C, appear to have been observed in a phase for which they show the turn-on of their OIR radiation.  Such a turn-on is presumably related to the unmodeled  relativistic electron acceleration at very early times and provides a test of explosion models.}
\item{None of the objects requires the inclusion of a uniformly distributed absorbing component to the early time radio absorption ($K_2$ in Equation 1 of \cite{Weiler01-KW}) that implies a highly filamentary or clumpy structure in the CSM with which the blastwave is interacting.}
\item{Estimates for the color excess, $E(B-V)$, obtained from the modeling imply that many of the objects suffer appreciable extinction at the redshift of their parent galaxies because fitted zero redshift values significantly exceed estimates of Galactic extinction in their directions.  The values for the excess extinction at the redshifts of the host galaxies noted above and in Table~\ref{tab2-KW} are rather typical for late-type galaxies.}
\item{All of the RGRBs except for GRB980425/SN1998bw have similar observed peak 6 cm radio spectral luminosities of $\sim10^{31}~{\rm ergs\ s^{-1}\ Hz^{-1}}$.  This implies that GRBs are rare objects for which one must survey very large volumes of space in order to detect examples.  Such examples are, therefore, generally at very large distances and we detect only the brightest members of the luminosity distribution.  GRB980425/SN1998bw is the only known nearby, lower luminosity exception.}
\item{The prevalence of interstellar scintillation (ISS) in the early radio emission from many of these objects implies that the RGRBs are initially very small in angular size although they expand rapidly.}
\end{enumerate}

\subsection{Discussion}

Although many useful conclusions can be drawn from the light curve parameterizations, one should keep in mind that a more detailed study of radio GRBs will have to take into account several physical effects that are not seen in RSNe, and that have not been included in this brief overview, but need further consideration.

\subsubsection{Interstellar Scintillation: \label{scintillation}}

Because their high radio luminosity and low absorption allows detection at great distance and quite early when they are still of very small angular size, GRB radio afterglows appear to be so compact as to exhibit ISS during the first few days or weeks of detectability.  This possibility  was proposed by Goodman \cite{Goodman97} for GRB970508 based on earlier work by Rickett \cite{Rickett70} for pulsars.  After consideration of the several regimes of strong, weak, refractive, and diffractive scattering, Goodman \cite{Goodman97} concluded that both diffractive and refractive scintillation are possible for radio afterglows and that observation of the effects of scintillation can place limits on their $\mu$as (micro-arcsecond) sizes at levels far too small to be resolved with VLBI.  

{\bf GRB970508} shows strong flux density fluctuations at both 8.46 and 4.86 GHz until age $\sim$4 weeks after explosion, that Waxman \etal~ \cite{Waxman98} attributed to diffractive scintillation.  After $\sim$1 month, they found that the  modulation amplitude decreased, which is consistent with the diffractive scintillation being quenched by the increased size of the radio emitting region.  They also took this increasing source size to be consistent with, and supportive of, the ``fireball'' model predictions (see $\S$ \ref{fireball}).  From their 4.86 and 8.46 GHz results, Waxman \etal \cite{Waxman98} concluded that the quenching of diffractive scintillation at $\sim$4 weeks implies a size at that epoch of $\sim 10^{17}$ cm and an expansion speed comparable to that of light.  

In contrast to the conclusion of Waxman \etal \cite{Waxman98} that GRB970508 is undergoing strong diffractive scintillation during the first 30 days after explosion, Smirnova \& Shishov \cite{Smirnova00} concluded that the radio afterglow is, in fact, undergoing only weak scintillation at 4.86 and 8.46 GHz at early times with refractive scintillation dominating at 1.43 GHz.  

{\bf GRB980329} shows rapid flux density fluctuations at 4.9 and 8.3 GHz that are extinguished by age $\sim$3 weeks.  Although they did not analyze the scintillation data in detail, Taylor \etal \cite{Taylor98b} pointed out a similarity to the better studied scintillations of GRB970508 and suggested that the early-time angular size of GRB980329 may be even smaller than the $\sim$ 3 $\mu$as inferred for GRB970508 by Goodman \cite{Goodman97} and Waxman \etal \cite{Waxman98} because GRB980329 is at a lower Galactic latitude, that should more quickly quench ISS.

{\bf GRB980519} shows strong modulation of the flux density at 4.86 and 8.46 GHz during the first $\sim$20 days after the GRB burst, that Frail \etal \cite{Frail00b} interpreted as being due to diffractive ISS.  They derived a resulting maximum radio source size of $<$ 0.4 $\mu$as, an extremely compact object.  As was seen for GRB970508, the 1.4 GHz emission from GRB980519 seems to be suppressed, in this case below their detectability limit because their three measurements at 1.4 GHz are all upper limits.  Frail \etal \cite{Frail00b} attributed this suppression to synchrotron selfabsorption (SSA) with a turn-over frequency between 1.43 and 4.86 GHz.  

\subsubsection{Cosmological Effects:}

Because the GRBs are at cosmological distances ($z \sim 1$), there are two effects that must be taken into account: 

\begin{enumerate}
\item{There is a time dilation that slows the light curve evolution in the observer's frame with respect to the time evolution in the emission frame.  This is a straightforward correction and has been elaborated by a number of authors (see, e.g., \cite{Colgate79,Hamuy93,Leibundgut90}).  The time dilation results in a true emitted time to 6 cm peak flux density [$(t_{\rm 6cm\ peak} - t_0)_{\rm emit}$] shorter than that actually observed [$(t_{\rm 6cm\ peak} - t_0)_{\rm obs}$].  The correction takes the form

\begin{equation}
(t_{\rm 6cm\ peak} - t_0)_{\rm emit} = (t_{\rm 6cm\ peak} - t_0)_{\rm obs}~(1 + z)^{-1}
\end{equation}
}

\noindent and must be applied to the measured times from explosion to 6 cm peak flux density to obtain true times.

\item{There is a correction of the observed flux density,  $(S_{\rm 6cm})_{\rm obs}$, owing to the redshift making the observed frequency different from that emitted, $(S_{\rm 6cm})_{\rm emit}$, for sources with non-zero spectral indices.  The correction normally takes the form, for $S \propto \nu^{+\alpha}$, 

\begin{equation}
(S_{\rm 6cm})_{\rm emit} = (S_{\rm 6cm})_{\rm obs}~(1 + z)^{-\alpha}
\end{equation}

\noindent However, Chevalier \& Li \cite{Chevalier00} have proposed an ``equality of peaks'' on theoretical grounds so that such a correction may be less important than expected.}
\end{enumerate}

\subsubsection{Relativistic Effects: \label{relativity}}

Essentially all researchers agree that the GRB phenomenon involves relativistic motion of the emitting region.  Whether the motion involves a spherical, relativistic fireball or a directed relativistic jet is probably not of concern from the radio observer's standpoint because the emission from the CSM interaction is probably not highly directed.  

For relativistic corrections there are two factors -- the relativistic motion factor $\Gamma$ [$\Gamma = ({1-\frac{v^2}{c^2}})^{-\frac{1}{2}}$] and, if the motion is directed, the viewing angle $\theta$.  If, as expected, the radio emission is not highly directed, then taking $\theta = 0$ leaves only the relativistic motion factor $\Gamma$ to be determined.

The factor $\Gamma$ is difficult to estimate from observations.  Theoretical modeling predicts $\Gamma$ of several hundred very early in the expansion phase \cite{Waxman97} declining to subrelativistic motion after only a few days to a few weeks \cite{Dai99,Huang98}.  However, although highly speculative, there is, perhaps, one method for estimating $\Gamma$.  

Weiler \etal \cite{Weiler01-KW} pointed out that by assuming a relation between Type Ib SN1998bw and GRB980425, they could obtain estimates of $\Gamma \sim$ 1.6 -- 2.0 for the radio afterglow.  They obtained one estimate by postulating that SN1998bw was expanding at $\sim$230,000 \kms if its brightness temperature at 6 cm peak flux density was $10^{11.5}$ K, as required by the inverse Compton limit.  The second estimate was obtained by assuming that the true 6 cm peak spectral luminosity was the same as that for the average of all known Type Ib/c RSNe.  Then, if one assumes: (a) all GRB radio afterglows arise in Type Ib/c SNe and (b) the postulate of Weiler \etal \cite{Weiler98-KW} that all Type Ib/c SNe may be approximate standard radio candles with an $L_{\rm 6cm\ peak} \sim1.3 \times 10^{27} \ {\rm erg}\ {\rm s}^{-1}\ {\rm Hz}^{-1}$, and (c) any increase in the observed flux density is solely due to relativistic boosting, then one can estimate a relativistic $\Gamma$ factor for each GRB radio afterglow.

Applying these assumptions to the peak 6 cm radio luminosity ($L_{\rm 6cm\ peak}$) values obtained for the GRBs and listed in Table~\ref{tab2-KW}, we obtain $\Gamma$ values $\sim10$. Such $\Gamma$s are similar to those found for VLBI ``superluminal'' sources (see e.g., \cite{Kellermann85,Scheuer84}) and for the Galactic micro-QSO superluminal sources  \cite{Hjellming95,Levinson96,Mirabel94,Tingay95}.  

\subsection{Theoretical Models \label{models}}

The origin of $\gamma$-ray bursts is a long standing problem dating back to the 1960s.  With little actual data to constrain them, many theories were developed.  Ruderman \cite{Ruderman75} first summarized these in the mid-1970s.  Cavallo \& Rees \cite{Cavallo78} were probably the first to use the term ``fireball'' in an article discussing several possibilities for the conversion of massive amounts of injection energy into $\gamma$-rays, including the possibility of relativistic expansion and the conversion of kinetic energy into $\gamma$-ray luminosity through the impact on an external medium.

The 1980s brought the more detailed calculations of Paczynski \cite{Paczynski86} and Goodman \cite{Goodman86} that suggested that the GRBs might be at cosmological distances of $z \sim 1$ or more and that the energies involved were at least comparable to that expected from an SN ($\sim 10^{51}$ ergs).  In the 1990s, of course, the field finally came of age with the \CO to provide almost daily detections of GRBs and \B to provide rapid followup for many bursts and accurate position determination for X-ray afterglows.  This increased data flow refined the model studies for the origin of GRBs so that there are basically only two models presently under serious consideration -- relativistic fireball and relativistic jet.  Because of the observational (particularly radio) orientation of this review, we shall not discuss the mechanisms for the source of the energy release, whether neutron star or black hole formation in SNe, coalescence of compact objects, or more exotic phenomena.  We shall be more concerned with the observable consequences of this great energy release and, where possible, similarities to known phenomena.

\subsubsection{Relativistic Fireball: \label{fireball}}

Many workers have concluded that the relativistic fireball model is the most likely scenario for GRB creation and most prefer conversion of the blastwave energy into emission through blastwave interaction with an external medium.  The basic outlines of the fireball model are probably best summarized by Waxman \cite{Waxman97}:  

\begin{quote}
A compact $r_0 \sim 10^7$ cm source releases an energy E comparable to that observed in $\gamma$-rays, E $\sim 10^{51}$ ergs, over a time t $<$ 100 s.  The large energy density in the source results in an optically thick plasma that expands and accelerates to relativistic velocity.  After an initial acceleration phase, the fireball energy is converted to proton kinetic energy.  A cold shell of thickness $ct$ is formed and continues to expand with time-independent Lorentz factor $\Gamma \sim$ 300.  The GRB is produced once the kinetic energy is dissipated at large radius, r $> 10^{13}$ cm, due to internal collisions within the ejecta and radiated as $\gamma$-rays through synchrotron and possibly inverse-Compton emission of blastwave accelerated electrons. Following internal collisions, that convert part of the energy to radiation and that result from variations in $\Gamma$ across the expanding shell, the fireball rapidly cools and continues to expand with approximately uniform Lorentz factor $\Gamma$.  As the cold shell expands, it drives a relativistic blastwave into the surrounding gas.
\end{quote}  

Huang \etal \cite{Huang98} described this interaction at larger radius as:

\begin{quote}
After producing the main GRB, the cooling fireball is expected to expand as a thin shell into the interstellar medium (ISM) and generate a relativistic blastwave, although whether the expansion is highly radiative \cite{Vietri97} or adiabatic is still controversial.  Afterglows at longer wavelengths are produced by the shocked ISM.
\end{quote}

Rees \& M{\'e}sz{\'a}ros \cite{Rees92-KW} described the possibility of generation of GRBs from the interaction of a highly relativistic ($\Gamma \sim 10^3$) blastwave interacting with a tenuous ISM or CSM to generate a $\sim 10^{51}$ erg $\gamma$-ray burst.  M{\'e}sz{\'a}ros \etal \cite{Meszaros93} worked this out in more detail (see also \cite{Meszaros95} for a review of fireball models).  When the first optical afterglow of a GRB was finally found for GRB970228, although the data were very limited Wijers \etal \cite{Wijers97-KW} found support their model of an expanding, relativistic fireball of total energy $\sim 10^{51}$ ergs plowing into a surrounding medium.

As early as 1993, Paczynski \& Rhoads \cite{Paczynski93} pointed out the similarity of GRB relativistic fireball interaction with CSM or ISM material to relativistic jets interacting with intergalactic material, supernova remnants (SNRs) interacting with the ISM, or RSNe interacting with the CSM and predicted the presence of delayed radio emission from the GRB process.  M{\'e}sz{\'a}ros \& Rees \cite{Meszaros97-KW} and Vietri \cite{Vietri97} also predicted late time radio, optical and soft X-ray emission based on the fireball model. 

Huang \etal~ \cite{Huang98} proposed that the extremely large $\Gamma$ factors of the GRB-producing relativistic fireball decay quickly, within 3 to 4 days, so that the blastwave/CSM interaction generating the synchrotron emission is only slightly relativistic with $2 < \Gamma < 5$.  Even more rapidly evolving is the fireball model of Dai \etal \cite{Dai99} who proposed that the fireball is only mildly relativistic after $\sim$3 hours and non-relativistic after a few days.

Waxman \cite{Waxman97} studied the optical and radio afterglow from GRB970508 and found agreement with the cosmological fireball model for a fireball energy of $\sim 10^{52}$ ergs and the blastwave expanding into an ambient medium with density $\sim 1~{\rm cm}^{-3}$.  Waxman \etal \cite{Waxman98} also compared the radio light curves of GRB970508 to the predictions of the Waxman \cite{Waxman97} fireball model and, although their fit is extremely poor after $\sim$4 weeks of age, they ascribed the deviation to transition of the fireball from relativistic to subrelativistic expansion with a Lorentz factor $\Gamma < 2$ or, possibly, to a lower energy ``jet'' structure for the fireball at early epochs.  Their model also predicts transition to a more uniformly emitting fireball structure by age $\sim$5 weeks, becoming a spherical, non-relativistic emitter by age $\sim$12 weeks.

\subsubsection{Relativistic Jets and Hypernovae:}

Support for the relativistic fireball model is not universal, however.  Dar \cite{Dar98} proposed an origin for GRBs and their longer wavelength afterglows in relativistic jets produced by mergers and/or accretion-induced collapse of compact stellar objects.  He felt that such a scenario can better solve the ``energy problem'' (where some estimates of relativistic fireball energies can range up to a rather high $> 10^{52}$ ergs), the short time variability of many GRBs, and other apparent deviations from the relativistic fireball models.  A somewhat different variant on the fireball model has been proposed by Paczynski \cite{Paczynski98-KW} who suggested that a ``hypernova'' explosion produces a ``dirty'' fireball from the violent death of a massive star.

\subsubsection{Multiple Origins:}

Livio \& Waxman \cite{Livio00-KW} took the broader stance that the GRBs with afterglows, although probably not originating in neutron star--neutron star (ns-ns) mergers, could well have several possible origins including the collapse of massive stars as in SNe (GRB980519 and GRB980326) or could be produced by Black Hole-He star mergers (GRB990123 and GRB990510), with a third group such as GRB970228 and GRB970508 being possibly produced by Black Hole-He star mergers but without collimated jets being formed by the collapse.  They disagreed with other workers that the transition from relativistic expansion to non-relativistic expansion of a spherical fireball will produce a detectable break in the afterglow decline rate.  They also felt that the GRBs creating detectable afterglows are diverse with only some producing highly collimated jets, but with all  originating from massive stars and all interacting with a stellar wind environment.

Perhaps the broadest summary and comparison of models with radio data for several GRBs previous to this review is that given by Chevalier \& Li \cite{Chevalier00} (see also, \cite{Chevalier99-KW}) who were able to describe rather well the radio light curves of several GRB radio afterglows as arising from blastwave interaction with an external medium.  They also suggested that there is evidence for two groupings:  (a) GRBs arising from the explosion of a massive star, possibly a Wolf-Rayet star, embedded in a dense $\rho \propto r^{-2}$ stellar wind-formed CSM and accompanied by a supernova, and (b) GRBs arising from the explosion caused by a compact binary merger that occur in a constant density ISM and are not accompanied by a supernova.

\section{Summary} 

We have assembled an overview of the radio emission from GRBs and have shown that there are many similarities to the much better understood radio emission from RSNe.  In particular, we have shown that the formalism described by Weiler \etal \cite{Weiler01-KW} can be applied to the radio and OIR emission from a number of GRBs and provides a satisfactory description of their light curves over extremely broad frequency and time ranges.  These results then imply that at least some GRBs probably arise in the explosions of massive stars in dense, highly structured circumstellar media.  Such media are presumed to have been established by mass-loss from the preexplosion stellar systems.  

\section{Acknowledgments}

KWW, \& MJM thank the Office of Naval Research (ONR) for the 6.1 funding supporting this research.  Additional information and data on radio emission from SNe and GRBs can be found on {\it http://rsd-www.nrl.navy.mil/7214/weiler/} and linked pages.

%

\end{document}